# An Open Source Testing Tool for Evaluating Handwriting Input Methods

Liquan Qiu, Lianwen Jin+, Ruifen Dai, Yuxiang Zhang, Lei Li
South China University of Technology
Guangzhou, China
czqiuliquan@gmail.com, +eelwjin@scut.edu.cn, 330167372@qq.com, 532069838@qq.com, eelilei@scut.edu.cn

*Abstract*—**This paper presents an open source tool for testing the recognition accuracy of Chinese handwriting input methods. The tool consists of two modules, namely the PC and Android mobile client. The PC client reads handwritten samples in the computer, and transfers them individually to the Android client in accordance with the socket communication protocol. After the Android client receives the data, it simulates the handwriting on screen of client device, and triggers the corresponding handwriting recognition method. The recognition accuracy is recorded by the Android client. We present the design principles and describe the implementation of the test platform. We construct several test datasets for evaluating different handwriting recognition systems, and conduct an objective and comprehensive test using six Chinese handwriting input methods with five datasets. The test results for the recognition accuracy are then compared and analyzed.**

## I. INTRODUCTION

With the popularization of smart phones, tablet PCs, and other intelligent terminals, the widespread use of touch screens and the development of handwriting recognition technologies mean that textual input is no longer limited to the keyboard. Indeed, handwritten input is becoming increasingly popular. In recent years, researchers in the field of handwritten character recognition have made significant progress [1-5]. In the ICDAR 2011 online Chinese handwriting recognition competition, the maximum recognition rate was 95.77%, and this figure increased to 97.39% in 2013 [6,7]. Some academic research has fed into practical applications in mobile devices, such as the development of input methods for Chinese handwriting. However, limitations in computing and storage resources mean that not all cutting-edge technologies can be applied to mobile devices.

Currently, mainstream Chinese input methods (IME) such as Google Pinyin, Baidu, Sogou, SCUT gPen, iFLY, and Hanvon provide handwriting input modules. Some offer both single-character recognition and input as well as overlapping handwriting and text-line input methods. Of all the input methods offered by the Chinese mobile phone market, 10.5% of users stated that they prefer handwriting input [8]. This figure is the second-highest after Pinyin (76.7%), and much higher than voice input (4.6%) and Chinese stroke input (5.1%) [8]. If we assume there are 500 million smartphone users in China [9], this indicates that around 50 million people would prefer to use handwriting input. Thus, the handwriting input method is a significant technique, and is an important feature of applications for mobile phones and devices.

The input methods mentioned above provide relatively mature handwriting input, but none of them specify the underlying technology. Some developers of handwriting input methods have declared their products to have very high recognition performance, but there is a lack of strict and rigorous test evidence. Indeed, it is difficult to objectively compare the real recognition performance of different handwriting IMEs because there is neither an open testing tool nor standard test data available to evaluate them. In view of this, we have designed a test tool based on Android and the MonkeyRunner Tool [10] to conduct a large-scale objective evaluation of different Chinese handwriting input methods. In this paper, we describe the design of this tool, as well as the test data, methods, and evaluation criteria used. In addition to analyze the recognition accuracy, we also compare aspects of each input method, including their packet size, ROM occupation, CPU usage, and memory usage during the handwriting process. An objective test of six mainstream Chinese IMEs is conducted, and comparative results for the recognition performance are presented and analyzed.

The source code for our test tool and the test data that we describe in this paper will be publicly available for research purposes[1].

[1] https://github.com/HCIILAB/IME_Test.git

## II. SYSTEM FRAMEWORK

The framework of the test tool is illustrated in Figure 1. The test tool is composed of two parts: a PC and a mobile client. The former includes the MonkeyRunner API [10], which is an automated test tool for Android device applications. The mobile client is designed and implemented based on the Android system. The test process proceeds as follows:

Step 1: The PC client loads the handwritten sample from a file and sends it to the mobile client.
Step 2: After the phone client has received the data, it simulates the handwriting process on the screen of the mobile phone.
Step 3: When a character sample has been completely written, the corresponding handwriting input method will be activated.
Step 4: The mobile client obtains and records the recognition result, and then calculates the recognition rate.

Step 5: Return to Step 1 if further test samples are available; otherwise, exit the program.

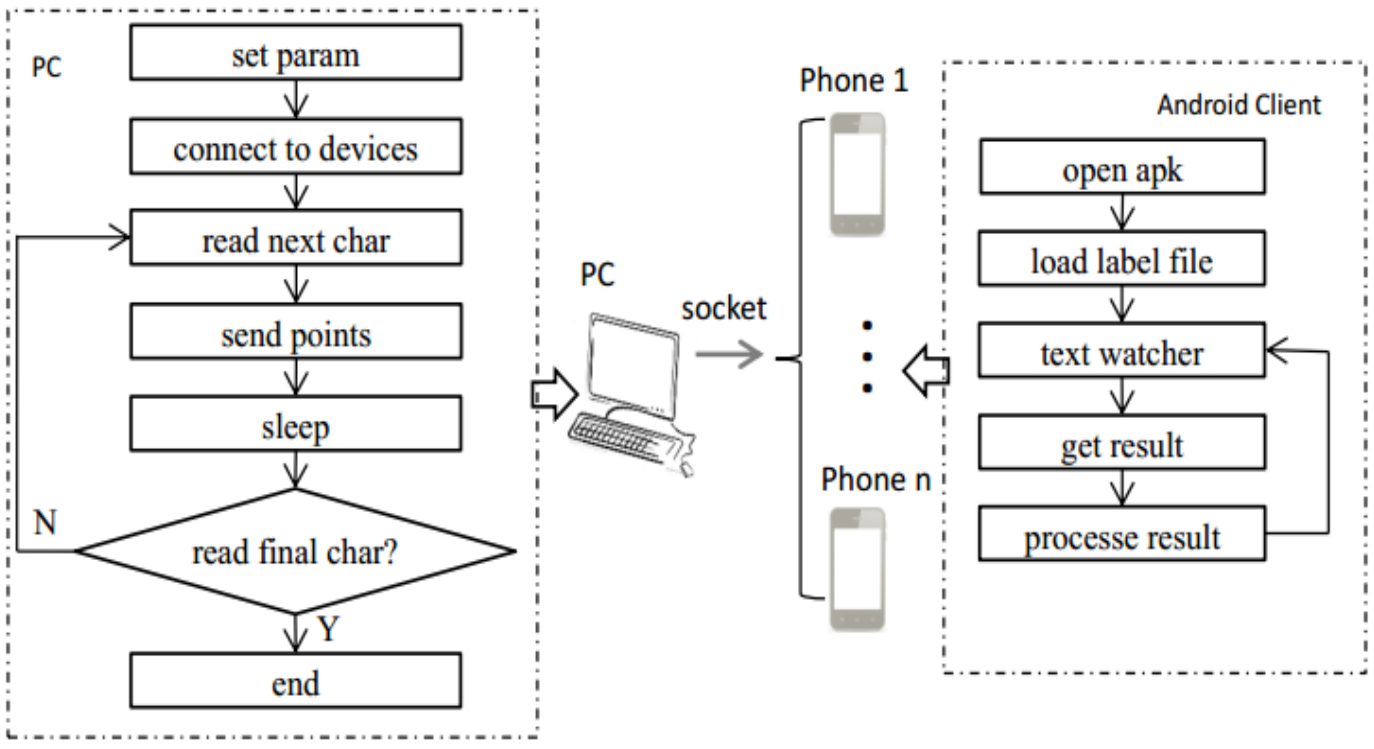

Figure 1 Testing tool framework

### A. PC client based on the MonkeyRunner tool

Handwriting samples are stored on the PC client. The sample data are read and sent to the mobile client automatically, and the mobile client then simulates the finger movements needed to write the sample on the screen. The mobile client then automatically tests the handwriting input methods on several devices at the same time. Google's MonkeyRunner tool meets the technical requirements of this step. MonkeyRunner can control one or more Android devices simultaneously, and provides an interface for sending touch events, enabling us to simulate the handwriting operation of a finger on the screen [10].

We have developed a Python script that runs on the PC client. This can read the trajectories of handwritten characters from the test sample file, and then sends the corresponding actions to the mobile device through the function interface provided by MonkeyRunner. The handwriting strokes are categorized into three types: pen-down points, pen-moving points, or pen-up points. Categorized touch/write events are sent to the Android client device, which is connected to the PC client via a USB under the socket communication protocol.

When simulating handwriting on the touch screen, the device will resample the handwriting trajectories so that the IME program receives the correct handwriting data. This process requires regular time intervals between points (denoted as t1). However, after the trajectory of a character has been sent, there is a short period of sleep time (denoted as t2) in which the IME identifies the character. For the same device, higher value of t1 results in a higher number of received handwriting points after resampling (but this will increase the processing time and the amount of data). The value of t2 must be greater than the waiting time in which the IME starts the recognition process and outputs its result (this value is usually greater than 300 ms).

### B. Android client

The main function of the Android client is to obtain the recognition results for the input method, and then compare these to the ground truth, save the recognition results, and compute the recognition accuracy.

As shown in Figure 2(a), the client has three main functional areas. The *TextView* area is used to display the current recognition accuracy. The *Button* area is used to load labeling files, and *EditText* is a text box that displays the recognition result for the IME. The client has additional monitoring and statistics modules. The former is used to monitor any change of status in the *EditText* area. Once the IME recognizes the input and submits this to *EditText*, the length of the text in the box will change. As the monitoring module obtains results, they are passed to the statistics module for comparison with the relevant ground truth label.

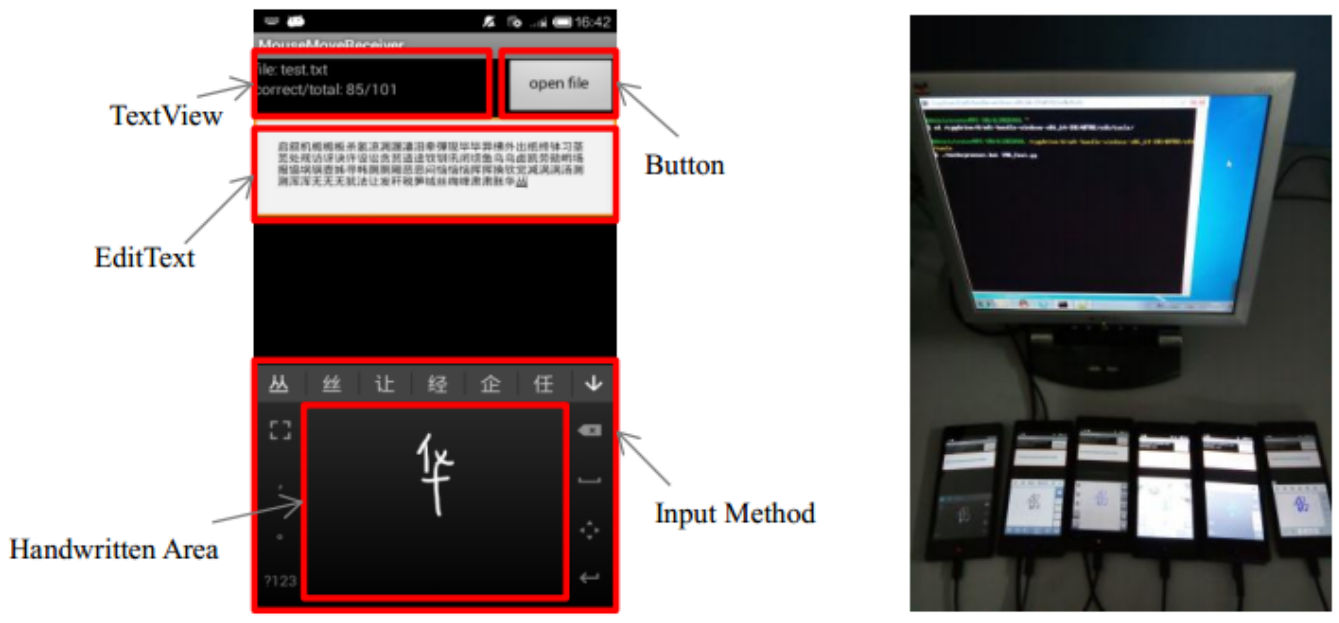

a) b)

Figure 2 Pictures of the test platform

## III. Preparation of the Test Dataset

Our test data is taken from the eight databases detailed in Table 1. By randomly sampling from these datasets, we constructed five test sets described in Table 2. These comprise a simplified Chinese set (denoted as SimpleChar) in GB2312-80 standard, traditional Chinese set (denoted as TradChar) in Big5 standard, mixed simplified and traditional Chinese set (denoted as SimpTradChar), rarely-used Chinese character set (denoted as RarelyUsedChar), and symbol set (denoted as SymbolChar). Note that SymbolChar contains uppercase and lowercase letters, numbers, punctuation, common symbols, and so on, as shown in Table 3.

The samples for TradChar, RarelyUsedChar, and SymbolChar were selected from In-House DB2; SimpleChar samples were selected from the other seven databases. SimpTradChar is a combined set formed of SimpleChar and TradChar. All test data are publicly available, and can be downloaded from the website of our laboratory [2].

Table 1 Sample databases

| | CASIA OLHWDB 1.0[1] | CASIA OLHWDB 1.1 [1] | CASIA OLHWDB 1.2 [1] | 863 [12] | SCUT-COUCH [13] | HKU [14] | In-House DB1 | In-House DB2 |
|---|---|---|---|---|---|---|---|---|
| Number of samples | 143600 | 98235 | 79154 | 40578 | 161166 | 20863 | 184089 | 994500 |

If we take the SimpleChar dataset as an example, we randomly selected 20,000 samples (without repetition) from the datasets listed in Table 1 to form a test set. Repeating this operation five times produced five SimpleChar datasets, which we denote as SimpleChar_1~SimpleChar_5 accordingly. The other test sets were constructed in the same manner. As there are fewer samples in SymbolChar, the number of test samples in this category was 10,000 for each subset.

[2] http://www.hcii-lab.net/data/onHCCTestDataset/onHCCTest.html

The SCUT gPen handwriting IME developed by SCUT-HCII Laboratory was used in the experiments. However, the test datasets described above were not used to train our handwriting recognition engine.

Table 2 Number of characters and samples in the five test sets

| Sets | SimpleChar | TradChar | SimpTradChar | RarelyUsedChar | SymbolChar |
|---|---|---|---|---|---|
| # of classes | 6763 | 5401 | 8817 | 785 | 196 |
| # of samples | 813288 | 540100 | 1353388 | 78500 | 19600 |

Table 3 SymbolChar Class and RarelyUsedChar Class

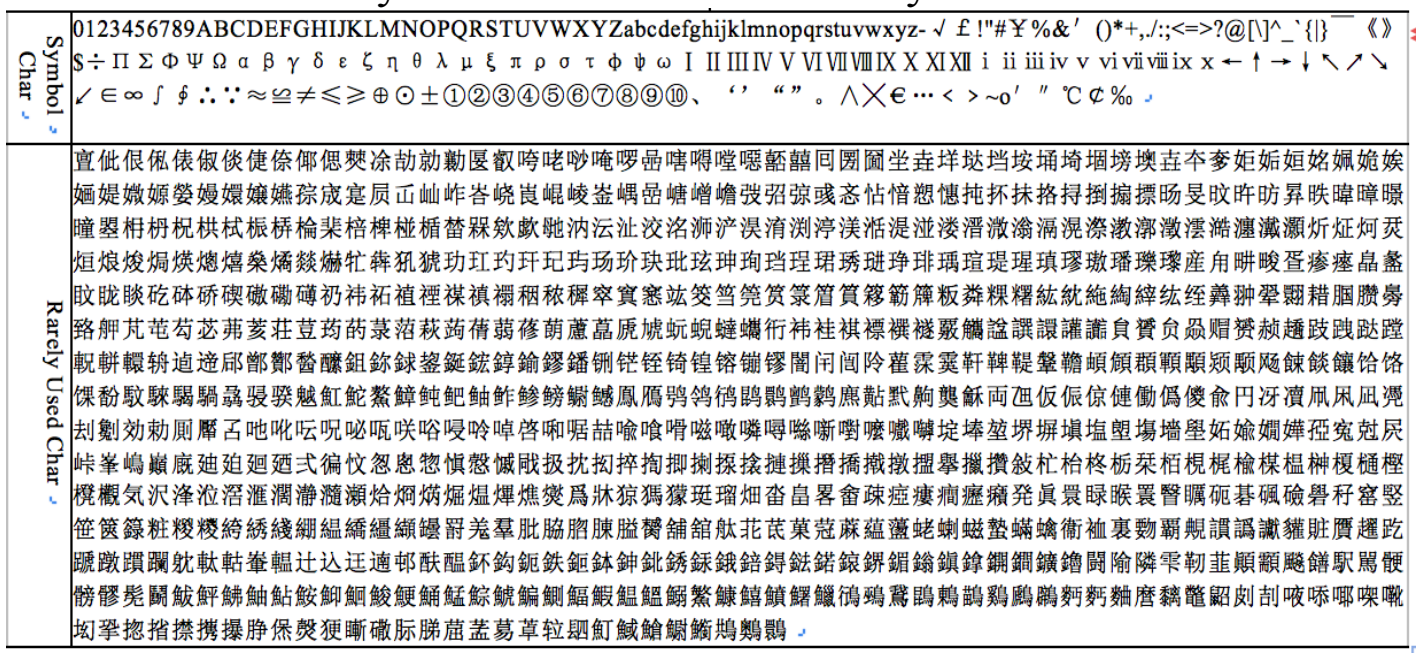

To conduct the tests, we used six Xiaomi Red Rice 1S Android phones [8], and set t1 = 0.006 s, t2 = 1.2 s.

## IV. EXPERIMENTAL RESULTS AND ANALYSIS

### *A. Experimental results*

We conducted automatic testing on several leading Chinese handwriting IMEs to objectively evaluate and compare their recognition performance. Currently, there are many Chinese IMEs in the market, including Baidu [16], Sogou [17], SCUT gPen [18], iFLY [19], Hanvon [20], Google Pinyin [21], TouchPal [22], and QQ [23]. We selected six of these for testing. To avoid the unintentional effects on the IME providers, the specific name of the IME will not be mentioned in later experiments. Instead, we will use the following symbols to denote them:

- BDD: Input Method 1
- gPen: Input Method 2
- GGG: Input Method 3
- XFF: Input Method 4
- SGG: Input Method 5
- HWW: Input Method 6

Table 4 gives some relevant information about several versions of the input method.

Table 4. Information of six input methods

| | BDD | SGG | gPen | XFF | GGG | HWW |
|---|---|---|---|---|---|---|
| Version | V5.1.5 | V7.0 | V4.0.0 | V5.0.1680 | V4.0.0 | V2.2.23 |
| Released date | 20140928 | 20141111 | 20141023 | 20141028 | 20141106 | 20120110 |

Before testing, we compared the package size, ROM occupation, CPU usage, and memory usage during handwriting of the IMEs. The results are shown in Figure 3.

As it can be seen from Figure 3, the installation packages and ROM sizes of each input method are different because different input methods use different handwriting recognition classifiers and dictionary models. In general, IMEs with more complex and larger classifier recognition dictionaries exhibit higher recognition accuracy but will require more memory space and higher CPU performance. Additionally, some input methods have a word association corpus and language model that requires extra storage space. BDD and HWW have smaller package and ROM sizes, SGG has a larger installation package and ROM size, and GGG has a higher memory usage and CPU usage. Finally, it should be pointed out that the installation package of BDD does not include a handwriting recognition engine. This must be downloaded from the Internet, which results in an additional file size of about 3 MB.

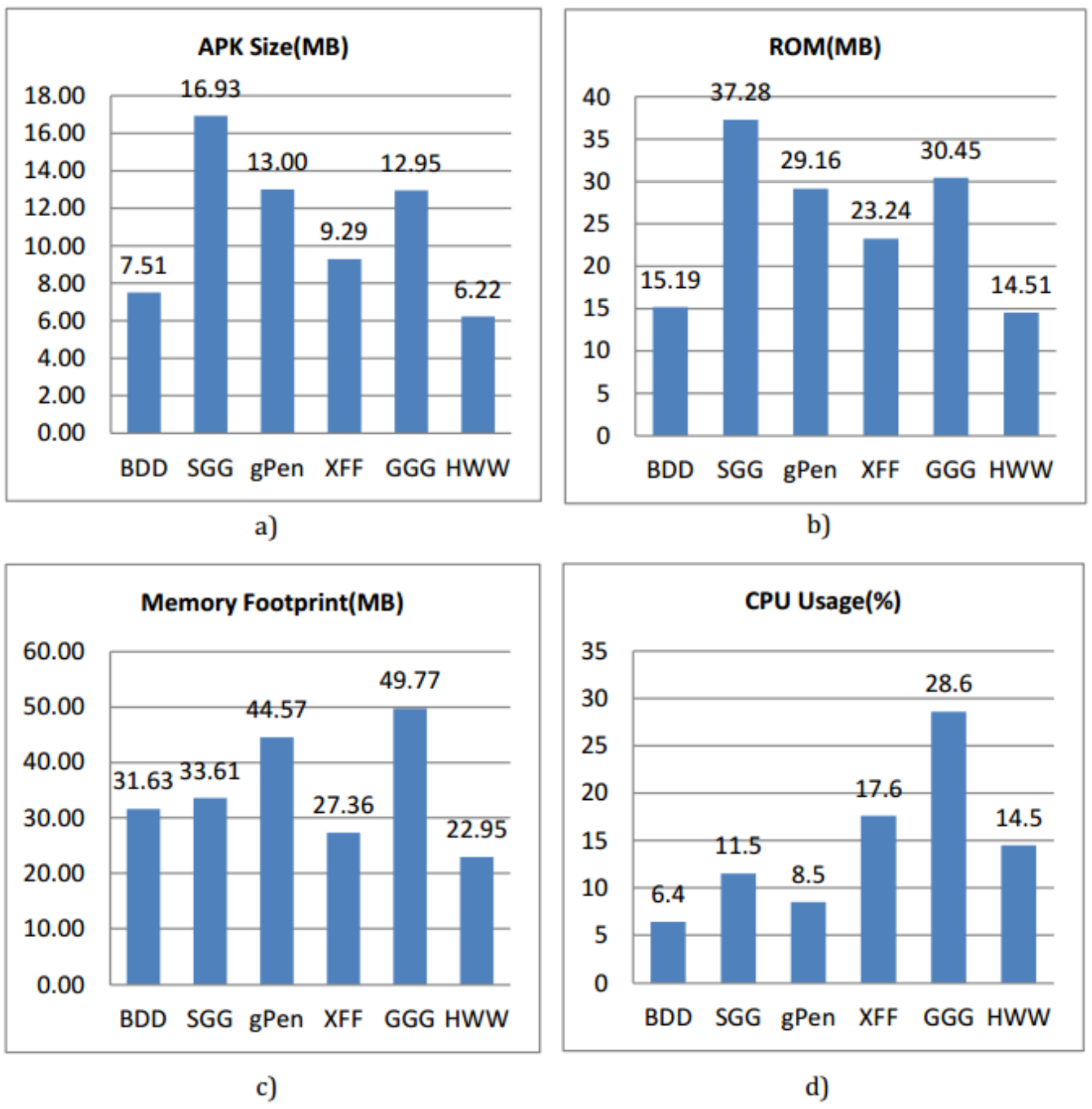

Figure 3 Performance of different IMEs.

Tables 5–9 present the recognition rates for each IME with the five different test datasets.

Table 5 SimpleChar test results

| | BDD | SGG | gPen | XFF | GGG | HWW |
|---|---|---|---|---|---|---|
| SimpleChar_1 | 72.43 | 88.33 | 95.48 | 89.07 | 80.22 | 92.11 |
| SimpleChar_2 | 72.43 | 87.34 | 95.31 | 90.68 | 80.99 | 92.55 |
| SimpleChar_3 | 71.56 | 87.84 | 95.45 | 90.30 | 79.58 | 91.96 |
| SimpleChar_4 | 71.15 | 87.62 | 95.08 | 89.34 | 79.36 | 92.20 |
| SimpleChar_5 | 72.94 | 88.50 | 95.40 | 89.65 | 80.20 | 92.48 |
| Average | 72.10 | 87.93 | 95.34 | 89.81 | 80.07 | 92.26 |

Table 6 TradChar test results

| | BDD | SGG | gPen | XFF | GGG | HWW |
|---|---|---|---|---|---|---|
| **TradChar_1** | 66.91 | 91.15 | 94.34 | 90.08 | 79.84 | 91.06 |
| **TradChar_2** | 68.46 | 91.22 | 95.99 | 89.40 | 78.40 | 90.93 |
| **TradChar_3** | 67.92 | 91.10 | 95.32 | 90.28 | 77.62 | 91.03 |
| **TradChar_4** | 66.43 | 90.94 | 95.68 | 89.67 | 78.65 | 91.00 |
| **TradChar_5** | 67.48 | 90.83 | 96.14 | 90.13 | 81.52 | 91.13 |
| **Average** | 67.44 | 91.05 | 95.49 | 89.91 | 79.21 | 91.03 |

Table 7 SimpTradChar test results

| | BDD | SGG | gPen | XFF | GGG | HWW |
|---|---|---|---|---|---|---|
| **SimpTradChar_1** | 71.08 | 90.15 | 96.48 | 90.39 | 78.95 | 91.02 |
| **SimpTradChar_2** | 70.65 | 89.31 | 94.85 | 89.94 | 78.68 | 91.82 |
| **SimpTradChar_3** | 70.86 | 89.47 | 96.41 | 90.33 | 81.36 | 91.39 |
| **SimpTradChar_4** | 70.38 | 89.60 | 95.31 | 89.84 | 81.76 | 90.98 |
| **SimpTradChar_5** | 70.08 | 89.77 | 96.29 | 90.18 | 78.97 | 90.71 |
| **Average** | 70.61 | 89.66 | 95.87 | 90.14 | 79.94 | 91.18 |

Table 8 RarelyUsedChar test results

| | BDD | SGG | gPen | XFF | GGG | HWW |
|---|---|---|---|---|---|---|
| **RarelyUsedChar_1** | 52.45 | 2.08 | 95.15 | 67.25 | 30.61 | 89.43 |
| **RarelyUsedChar_2** | 51.28 | 2.17 | 94.68 | 68.75 | 32.49 | 90.08 |
| **RarelyUsedChar_3** | 51.35 | 2.14 | 94.87 | 67.53 | 32.76 | 89.79 |
| **RarelyUsedChar_4** | 51.23 | 2.07 | 94.67 | 68.65 | 29.63 | 90.03 |
| **RarelyUsedChar_5** | 51.79 | 2.06 | 94.81 | 69.10 | 28.48 | 90.17 |
| **Average** | 51.62 | 2.10 | 94.84 | 68.26 | 30.79 | 89.90 |

Table 9 SymbolChar test results

| | BDD | SGG | gPen | XFF | GGG | HWW |
|---|---|---|---|---|---|---|
| **SymbolChar_1** | 42.80 | 56.19 | 84.71 | 20.39 | 30.82 | N/A |
| **SymbolChar_2** | 43.49 | 56.69 | 84.74 | 21.89 | 31.13 | N/A |
| **SymbolChar_3** | 43.77 | 56.43 | 84.96 | 22.35 | 30.60 | N/A |
| **SymbolChar_4** | 43.73 | 56.46 | 84.50 | 20.22 | 31.39 | N/A |
| **SymbolChar_5** | 43.25 | 56.51 | 84.40 | 19.98 | 30.76 | N/A |
| **Average** | 43.41 | 56.46 | 84.66 | 20.97 | 30.94 | N/A |

Table 10 Test results for SimpTradChar in overlap mode

| | BDD | gPen | XFF | GGG |
|---|---|---|---|---|
| **SimpTradChar_1** | 65.89 | 88.80 | 88.36 | 78.95 |
| **SimpTradChar_2** | 65.30 | 88.89 | 87.81 | 78.68 |
| **SimpTradChar_3** | 66.32 | 88.62 | 87.48 | 81.36 |
| **Average** | 65.84 | 88.77 | 87.88 | 79.66 |

Table 11 Test results for SimpTradChar in text-line mode

| | BDD | gPen | XFF | GGG |
|---|---|---|---|---|
| **SimpTradChar_1** | 60.10 | 81.38 | 89.80 | 78.95 |
| **SimpTradChar_2** | 60.20 | 80.43 | 90.45 | 78.68 |
| **SimpTradChar_3** | 60.52 | 81.52 | 89.97 | 81.36 |
| **Average** | 60.27 | 81.11 | 90.07 | 79.66 |

Two new methods of input, the text-line and overlap multi-characters modes, greatly improve the efficiency of text input. Single-character recognition is usually supported in both modes, but this can lead to lower recognition accuracy because the recognition engine needs to segment the character strings. Because segmentation algorithms have a certain error rate, the accuracy of the text-line IME naturally declined. We tested the recognition rate of the handwriting IMEs that support text-line and overlap multi-characters input modes. Table 10 gives the results for the overlap mode, and Table 11 presents the results for the text-line input mode.

*B. Analysis of test results*

Analyzing the results in Tables 5–9, we can make the following observations and conclusions:

(1) gPen IME produces the best recognition performance, with a significantly better recognition rate for every test dataset than other IMEs .

(2) With SimpleChar, TradChar, and SimpTradChar, the performance of SGG, XFF, and HWW are fairly good, whereas HWW is slightly higher than the other two.

(3) With RarelyUsedChar, gPen and HWW can recognize most characters with high accuracy, but GGG can only recognize a small proportion, and SGG produced hardly any correct results.

(4) The recognition rate of the alphanumeric and symbol sets is dramatically lower than with the other datasets for all IMEs except gPen. The test data contain a number of symbols that are rarely used on a phone, such as ∫, ∮, φ, ψ, √, ‰, and π. Most of these symbols are only supported by gPen. In addition, there are many similar characters in this dataset, such as x and X, o, O, and 0, 1 and l, and so on. For the HWW handwriting IME to recognize symbols, they must be written in a specific pre-defined region, otherwise the handwritten characters will not be recognized correctly.

(5) GGG does not achieve the high performance level as we expected. One reason may be that it integrates three input modes (single/text-line/overlap character input), causing the misrecognition of a single character as several characters, especially for Chinese characters with a left–right structure.

(6) The overall recognition rate of BDD is much lower than we expected, which is somehow contrary to our experience using mobile phones. We have not yet determined the cause of this problem. One possibility is that BDD may re-sample the input data. Re-sampling is often distance-based, i.e., when the value of the distance between a stroke point and the previous point is smaller than a certain threshold, the point is discarded. When we simulated handwriting input during the experiment, the size of the character was normalized to 180, which is smaller than the normal character size that is input by a user. This normalization leads to a reduction in distance between points, which may cause the input method to receive fewer stroke point data. Another resampling method is based on time, i.e., when the time interval between a stroke point and the previous point is below a certain threshold, the point will be discarded. When we simulated handwriting on the phones, t1 was set to 0.006 s. It is uncertain whether this setting prevented BDD from obtaining a sufficient number of points.

We experimented with different t1 settings, but BDD still did not produce particularly good performance.

From Tables 10 and 11, it can be seen that the text-line and overlap multi-characters input modes cause the accuracy of three IMEs to decrease, with only GGG producing exactly the same results. gPen suffers the highest decrease in accuracy, falling by 7.10% in overlap multi-characters mode and 14.76% in text-line mode. The accuracy of XFF decreases only slightly, by 2.26% and 0.07%, demonstrating the robustness of this IME in such modes.

Overall, it is clear that each IME has its own advantages. In terms of character recognition, gPen performs best, significantly better than the other handwriting IMEs. HWW, SGG, and XFF perform fairly well. For the recognition of rarely used characters, gPen and HWW outperform the other IMEs. In terms of symbol recognition, gPen is again the best, whereas for text-line and overlap multi-characters input modes, XFF displays impressive performance.

Note that the testing and statistics reported in this paper are based on the automatic simulation of handwriting on a touch-screen phone. Some errors may have been introduced by hardware limitations, communication errors, data sampling errors, and so on. We observed a slight gap between the test results and the real recognition rates of the input method developed by the authors' laboratory, where an average error rate of approximately 0.5–2% was observed. Although this test tool does not guarantee that the results will be strictly consistent with the real writing results, it is clearly a valuable and useful tool for estimating and testing the recognition accuracy of different IMEs.

## V. Conclusion

This paper presented an open source test tool for Chinese handwriting IMEs. By analyzing the performance of different IMEs in different situations, we demonstrated the feasibility and effectiveness of this tool.

During the experiments, we encountered problems such as differences between the simulated and real handwriting, and the t1 parameter setting. In addition, the proposed tool could not provide statistics for the recognition rate of the Top N ($N \geq 2$), and could not analyze the recognition rates of handwritten text lines and overlapping handwriting. These limitations remain open issues for further study.

## Acknowledgement

This research is supported in part by NSFC (Grant No.: 61472144), Research Fund for the Doctoral Program of Higher Education of China (Grant No.: 20120172110023).